\documentclass[twocolumn,aps,prd,superscriptaddress,floatfix,longbibliography]{revtex4-1}  
\usepackage{xcolor}
\usepackage[T1]{fontenc}
\usepackage[utf8]{inputenc}

\usepackage[draft]{changes}
\usepackage[normalem]{ulem}

\usepackage{graphicx}
\usepackage{amssymb} 
\usepackage{dcolumn}
\usepackage{bm}
\usepackage[mathlines]{lineno}
\usepackage[colorlinks,linkcolor=blue,anchorcolor=blue,citecolor=blue,urlcolor=blue]{hyperref}
\usepackage{multirow}
\usepackage{color}
\usepackage{amsmath}

\makeatletter
\renewcommand*{\@fnsymbol}[1]{\ensuremath{\ifcase#1\or \dagger\or *\or \ddagger\or
\mathsection\or \mathparagraph\or \|\or **\or \dagger\dagger \or
\ddagger\ddagger \else\@ctrerr\fi}} \makeatother

\usepackage{color}%

\usepackage[draft]{changes}
\usepackage[normalem]{ulem}
\definechangesauthor[name={gao}, color=blue]{gao}

\usepackage{color}%

\begin{document}

\title{Four-phonon scattering and coherent heat transport in ultrawide-bandgap SrSnO$_3$}

\author{Xuejie Li\footnotemark[1]}
\thanks{These authors contributed equally to this work.}
\affiliation{State Key Laboratory for Mechanical Behavior of Materials, State Key Laboratory of Porous Metal Materials, School of Materials Science and Engineering, Xi'an Jiaotong University, Xi'an 710049, China}

\author{Xiaoying Wang\footnotemark[1]}
\thanks{These authors contributed equally to this work.}
\affiliation{State Key Laboratory for Mechanical Behavior of Materials, State Key Laboratory of Porous Metal Materials, School of Materials Science and Engineering, Xi'an Jiaotong University, Xi'an 710049, China}

\author{Shengying Yue}
\affiliation{Laboratory for Multiscale Mechanics and Medical Science, State Key Laboratory for Strength and Vibration of Mechanical Structures, School of Aerospace, Xi'an Jiaotong University, Xi'an 710049, China}

\author{Turab Lookman}
\affiliation{State Key Laboratory for Mechanical Behavior of Materials, State Key Laboratory of Porous Metal Materials, School of Materials Science and Engineering, Xi'an Jiaotong University, Xi'an 710049, China}
\affiliation{AiMaterials Research LLC, Santa Fe, NM 87501, United States}

\author{Xiangdong Ding}
\affiliation{State Key Laboratory for Mechanical Behavior of Materials, State Key Laboratory of Porous Metal Materials, School of Materials Science and Engineering, Xi'an Jiaotong University, Xi'an 710049, China}

\author{Jun Sun}
\affiliation{State Key Laboratory for Mechanical Behavior of Materials, State Key Laboratory of Porous Metal Materials, School of Materials Science and Engineering, Xi'an Jiaotong University, Xi'an 710049, China}

\author{Zhibin Gao(Corresponding author)}
\email[E-mail: ]{zhibin.gao@xjtu.edu.cn}
\affiliation{State Key Laboratory for Mechanical Behavior of Materials, State Key Laboratory of Porous Metal Materials, School of Materials Science and Engineering, Xi'an Jiaotong University, Xi'an 710049, China}

\date{\today}

\begin{abstract}

SrSnO$_3$ is a promising ultrawide-bandgap perovskite oxide whose thermal transport is governed by structural distortions and anharmonic lattice dynamics. Here, we investigate the lattice thermal conductivity ($\kappa_L$) of orthorhombic and cubic SrSnO$_3$ within a unified first-principles framework combining self-consistent phonon renormalization, three- and four-phonon scattering, and coherent heat transport. Bonding analysis reveals a rigid Sn--O octahedral framework embedded in a weakly bonded Sr sublattice, giving rise to low-frequency vibrational modes susceptible to strong anharmonic effects. Four-phonon scattering is identified as a key mechanism limiting particle-like heat conduction, reducing the Peierls thermal conductivity by 19.4\% at 300 K in the orthorhombic phase and by 52.1\% at 1300 K in the cubic phase, while the coherent contribution provides a finite channel that partially compensates this reduction. We further show that the apparent agreement between three-phonon calculations and experimental thermal conductivity at room temperature is not indicative of a complete physical description. Instead, it arises from a near cancellation between four-phonon suppression of the particle-like channel and the neglected coherent contribution. This cancellation breaks down when the full temperature dependence is considered, where only the combined treatment improves agreement with the experimentally observed scaling behavior. A physically consistent description of $\kappa_L$ therefore requires phonon renormalization, four-phonon scattering, and coherent transport to be treated on equal footing rather than inferred from three-phonon agreement at a single temperature. These results provide microscopic insight into thermal transport in ultrawide-bandgap stannate perovskites and establish a benchmark for anharmonic transport in strongly distorted oxides.

\end{abstract}

\maketitle

\begin{figure*}[t!]
\centering
\includegraphics[height=0.72\textheight,keepaspectratio]{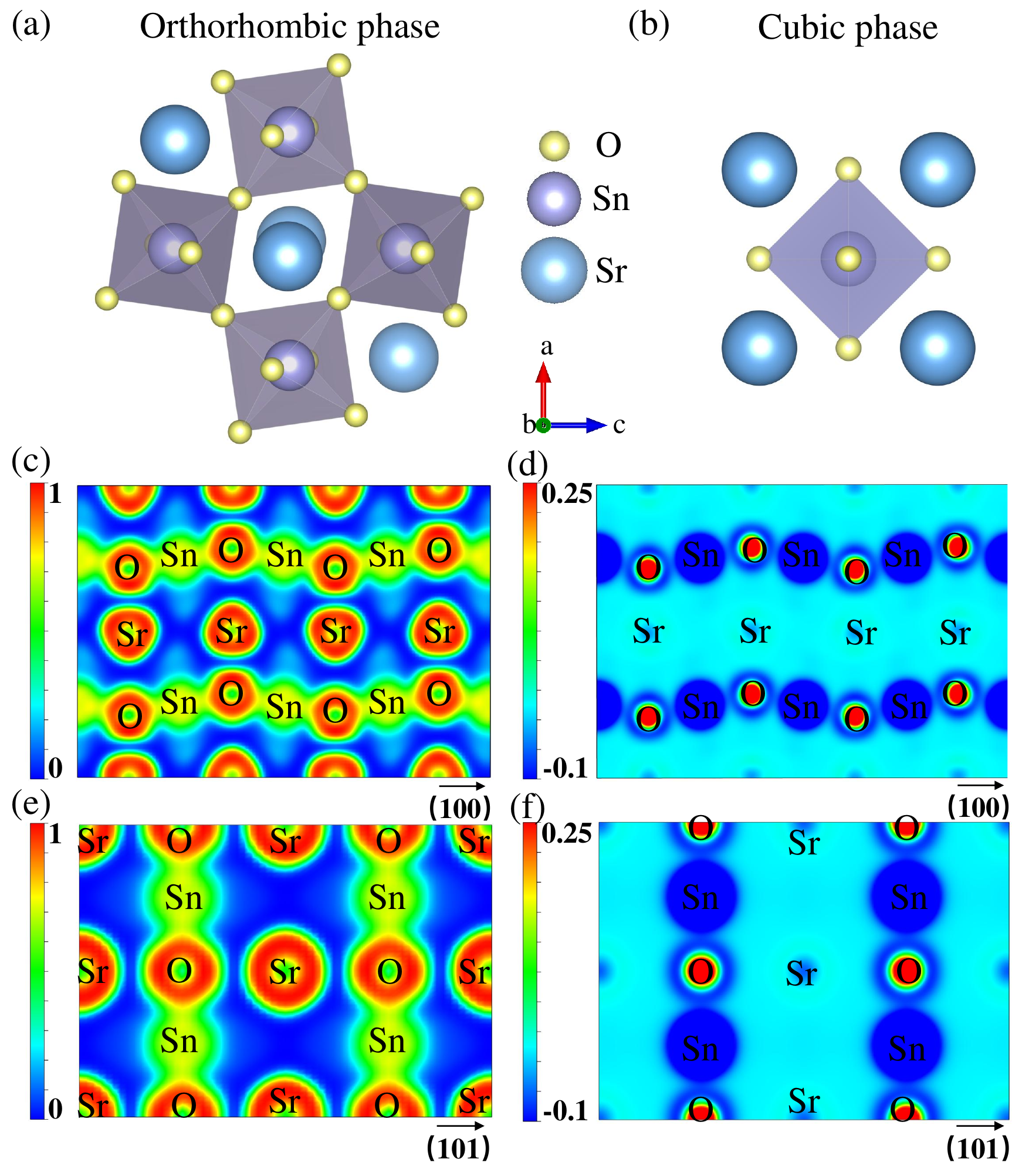}
\caption{
Crystal structures and real-space bonding analysis of SrSnO$_3$. (a,b) Orthorhombic $Pnma$ and cubic $Pm\bar{3}m$ structures, showing tilted and untilted SnO$_6$ octahedra, respectively. (c,e) Electron localization function maps on the (100) and (101) planes. (d,f) Charge-density-difference maps on the same planes. Both analyses consistently show stronger electron localization and charge redistribution along Sn-O bonds than around Sr-centered regions, establishing a rigid Sn–O backbone embedded in a weakly bonded Sr sublattice.
}
\label{fig1}
\end{figure*}

\begin{figure*}
\centering
\includegraphics[width=1\textwidth]{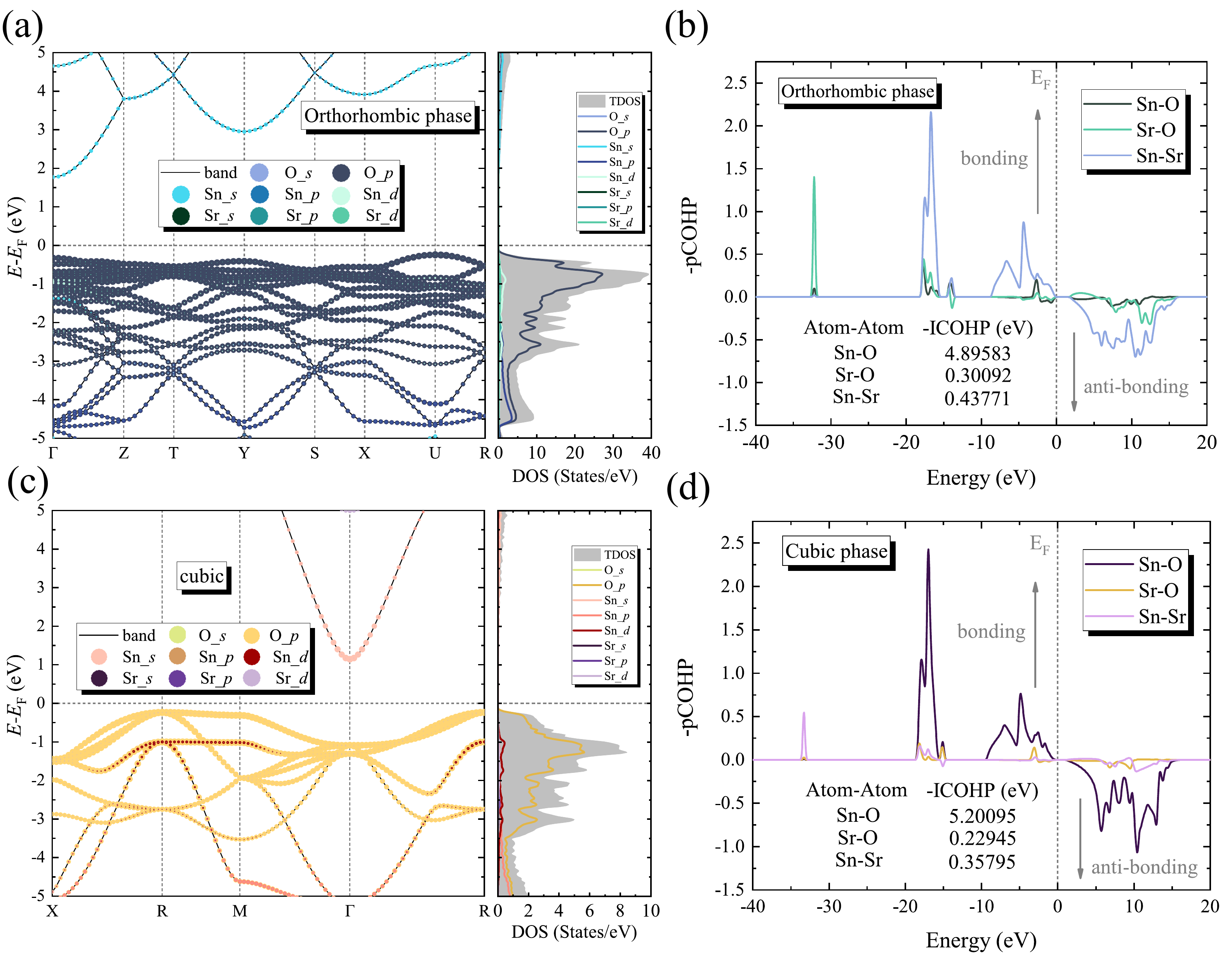}
\caption{
Orbital-resolved electronic structure and bonding analysis of SrSnO$_3$. (a,c) Projected band structures and density of states for the orthorhombic and cubic phases, showing O-$p$-dominated valence bands and Sn-$s$-dominated conduction-band edges with negligible Sr contribution near the band edges. (b,d) $-$ICOHP curves for selected atomic pairs calculated using LOBSTER. The integrated $-$ICOHP values confirm dominant Sn-O bonding (4.896 eV orthorhombic, 5.201 eV cubic) and much weaker Sr–O and Sn–Sr interactions, establishing a bonding hierarchy that underlies the low-frequency vibrational channels sensitive to anharmonicity. The zero of energy is set at the Fermi level.
}
\label{fig2}
\end{figure*}

\begin{figure}
\centering
\includegraphics[width=1.0\columnwidth]{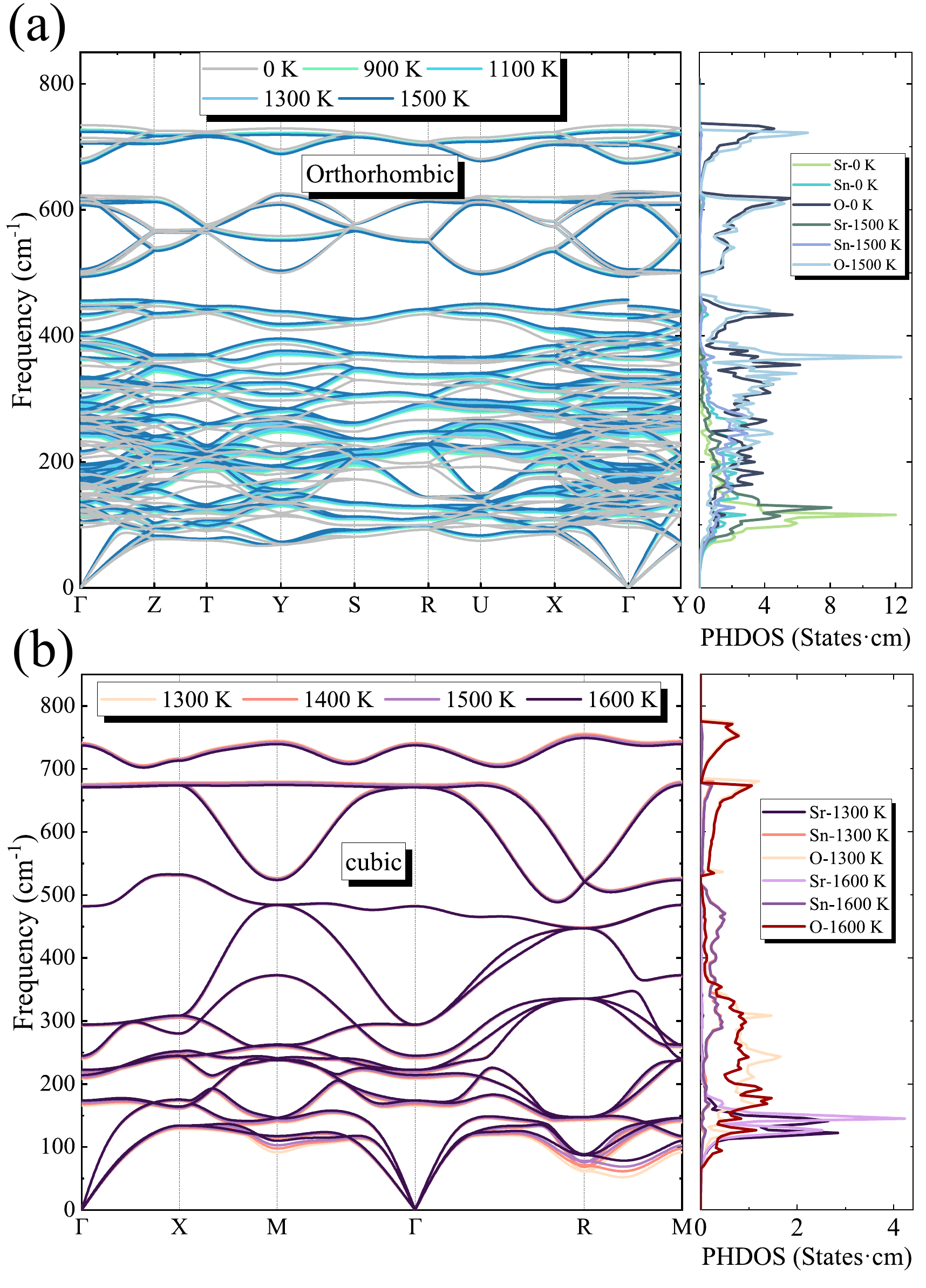}
\caption{
Temperature-dependent phonon dispersions and projected phonon density of states of SrSnO$_3$ in the (a) orthorhombic and (b) cubic phases, obtained from self-consistent phonon renormalization. The orthorhombic spectra are shown from 0 to 1500 K and the cubic spectra from 1300 to 1600 K. Low-lying optical branches below 250 cm$^{-1}$, associated with Sr displacements and octahedral framework motion, soften and shift with temperature, most strongly in the cubic phase; the high-frequency region is dominated by O vibrations. These renormalized low-frequency branches lie within the principal heat-carrying window and underlie both the 4ph scattering and the coherent contribution analyzed below.
}
\label{fig3}
\end{figure}

\begin{figure*}
\centering
\includegraphics[width=1.5\columnwidth]{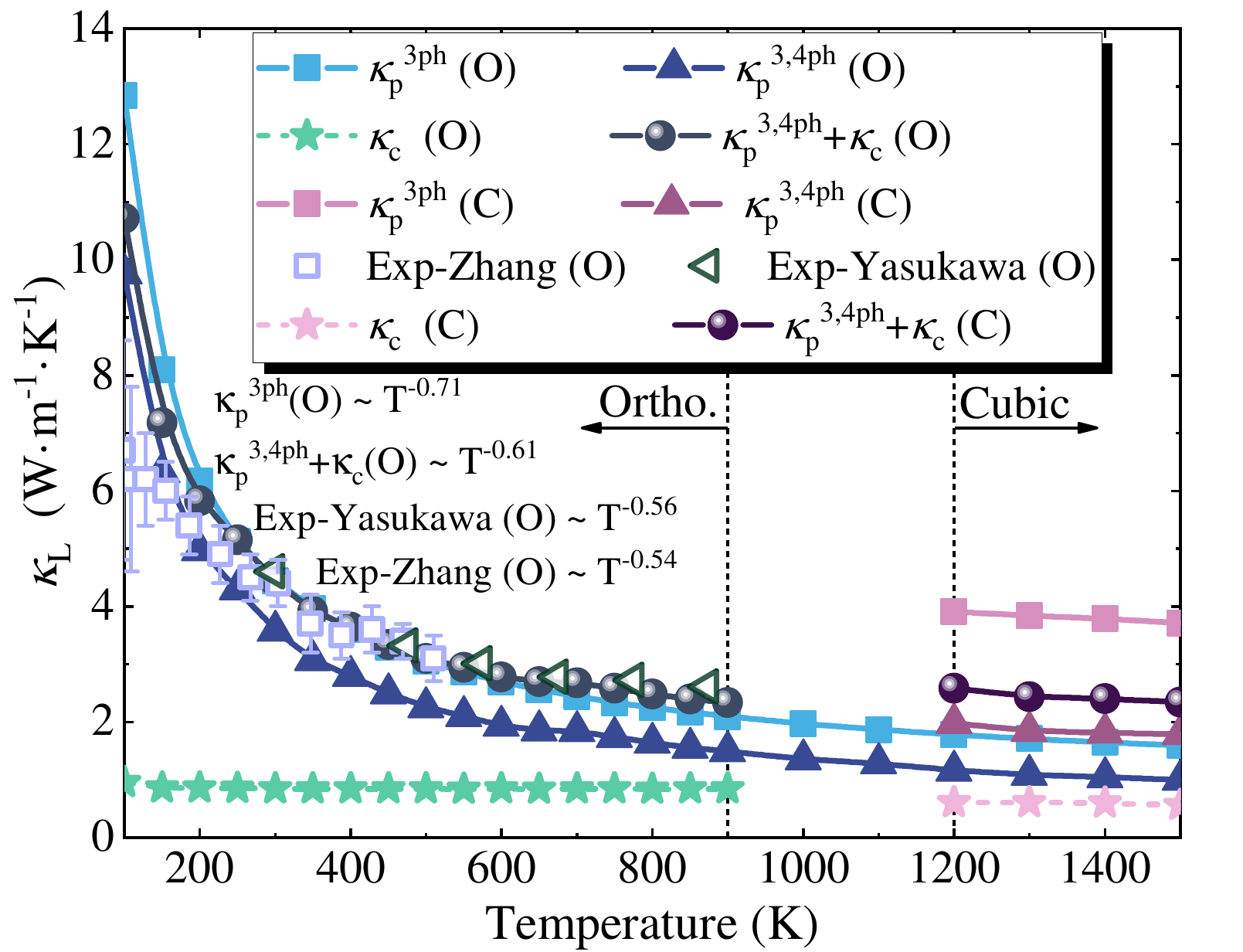}
\caption{
Temperature-dependent lattice thermal conductivity of orthorhombic and cubic SrSnO$_3$. The particle-like Peierls contributions calculated within the three-phonon (3ph) and combined three- and four-phonon (3,4ph) scattering channels are denoted as $\kappa_{\mathrm{p}}^{3\mathrm{ph}}$ and $\kappa_{\mathrm{p}}^{3,4\mathrm{ph}}$, respectively. The coherent contribution is labeled as $\kappa_{\mathrm{c}}$, and the total lattice thermal conductivity is given by $\kappa_{\mathrm{p}}^{3,4\mathrm{ph}}+\kappa_{\mathrm{c}}$. Experimental data from Zhang \textit{et al.} (epitaxial thin films) and Yasukawa \textit{et al.} (ceramics) are included for comparison. In the orthorhombic phase, the effective temperature dependence of the three-phonon contribution follows approximately $T^{-0.71}$, while inclusion of four-phonon scattering and coherent transport yields a weaker dependence of $T^{-0.61}$. Both experimental datasets exhibit a consistent scaling of approximately $T^{-0.54\sim -0.56}$, indicating that higher-order anharmonicity and coherent transport are essential to improve agreement with the observed temperature dependence. Cubic results below 1295 K are symmetry-constrained idealizations rather than equilibrium-phase predictions.
}
\label{fig4}
\end{figure*}

\begin{table}[htbp]
\centering
\small
\setlength{\tabcolsep}{3.8pt}
\renewcommand{\arraystretch}{1.08}

\caption{
Calculated lattice thermal conductivity (in W m$^{-1}$ K$^{-1}$) of orthorhombic (O) and cubic (C) SrSnO$_3$ at representative temperatures. $\kappa_{\rm p}^{3{\rm ph}}$ and $\kappa_{\rm p}^{3,4{\rm ph}}$ denote the particle-like Peierls thermal conductivity with three-phonon scattering and with combined three- and four-phonon scattering, respectively; $\kappa_{\rm c}$ is the coherent contribution and $\kappa_{\rm L}=\kappa_{\rm p}^{3,4{\rm ph}}+\kappa_{\rm c}$ is the total lattice thermal conductivity. The last column reports the four-phonon suppression of the particle-like channel, defined as $1-\kappa_{\rm p}^{3,4{\rm ph}}/\kappa_{\rm p}^{3{\rm ph}}$. The 1200 K cubic entry is an idealized symmetry-constrained value; the equilibrium phase is tetragonal.
}
\label{tab:kappa}

\resizebox{\columnwidth}{!}{%
\begin{tabular}{c c c c c c c}

\hline\hline
$T$ (K) & Phase &
$\kappa_{\rm p}^{3{\rm ph}}$ &
$\kappa_{\rm p}^{3,4{\rm ph}}$ &
$\kappa_{\rm c}$ &
$\kappa_{\rm L}$ &
$1-\kappa_{\rm p}^{3,4{\rm ph}}/\kappa_{\rm p}^{3{\rm ph}}$ \\
\hline

100  & O & 12.84 & 9.77 & 0.95 & 10.71 & 0.238 \\
150  & O & 8.10  & 6.32 & 0.86 & 7.18  & 0.220 \\
200  & O & 6.17  & 4.97 & 0.85 & 5.82  & 0.195 \\
250  & O & 5.12  & 4.30 & 0.85 & 5.15  & 0.160 \\
300  & O & 4.44  & 3.58 & 0.85 & 4.43  & 0.194 \\
350  & O & 3.96  & 3.07 & 0.84 & 3.91  & 0.226 \\
400  & O & 3.59  & 2.80 & 0.84 & 3.64  & 0.220 \\
450  & O & 3.29  & 2.50 & 0.84 & 3.34  & 0.239 \\
500  & O & 3.06  & 2.26 & 0.84 & 3.10  & 0.261 \\
550  & O & 2.87  & 2.10 & 0.84 & 2.94  & 0.268 \\
600  & O & 2.69  & 1.93 & 0.84 & 2.78  & 0.283 \\
650  & O & 2.56  & 1.85 & 0.84 & 2.69  & 0.277 \\
700  & O & 2.44  & 1.84 & 0.84 & 2.68  & 0.246 \\
750  & O & 2.34  & 1.74 & 0.84 & 2.58  & 0.256 \\
800  & O & 2.25  & 1.64 & 0.84 & 2.48  & 0.271 \\
850  & O & 2.17  & 1.55 & 0.84 & 2.39  & 0.285 \\
900  & O & 2.09  & 1.49 & 0.84 & 2.33  & 0.287 \\

1200 & C & 3.91 & 1.98 & 0.60 & 2.58 & 0.493 \\
1300 & C & 3.84 & 1.84 & 0.60 & 2.44 & 0.521 \\
1400 & C & 3.78 & 1.81 & 0.59 & 2.40 & 0.522 \\
1500 & C & 3.71 & 1.79 & 0.55 & 2.34 & 0.517 \\

\hline\hline
\end{tabular}%
}
\end{table}

\begin{figure*}
\centering
\includegraphics[width=0.96\textwidth]{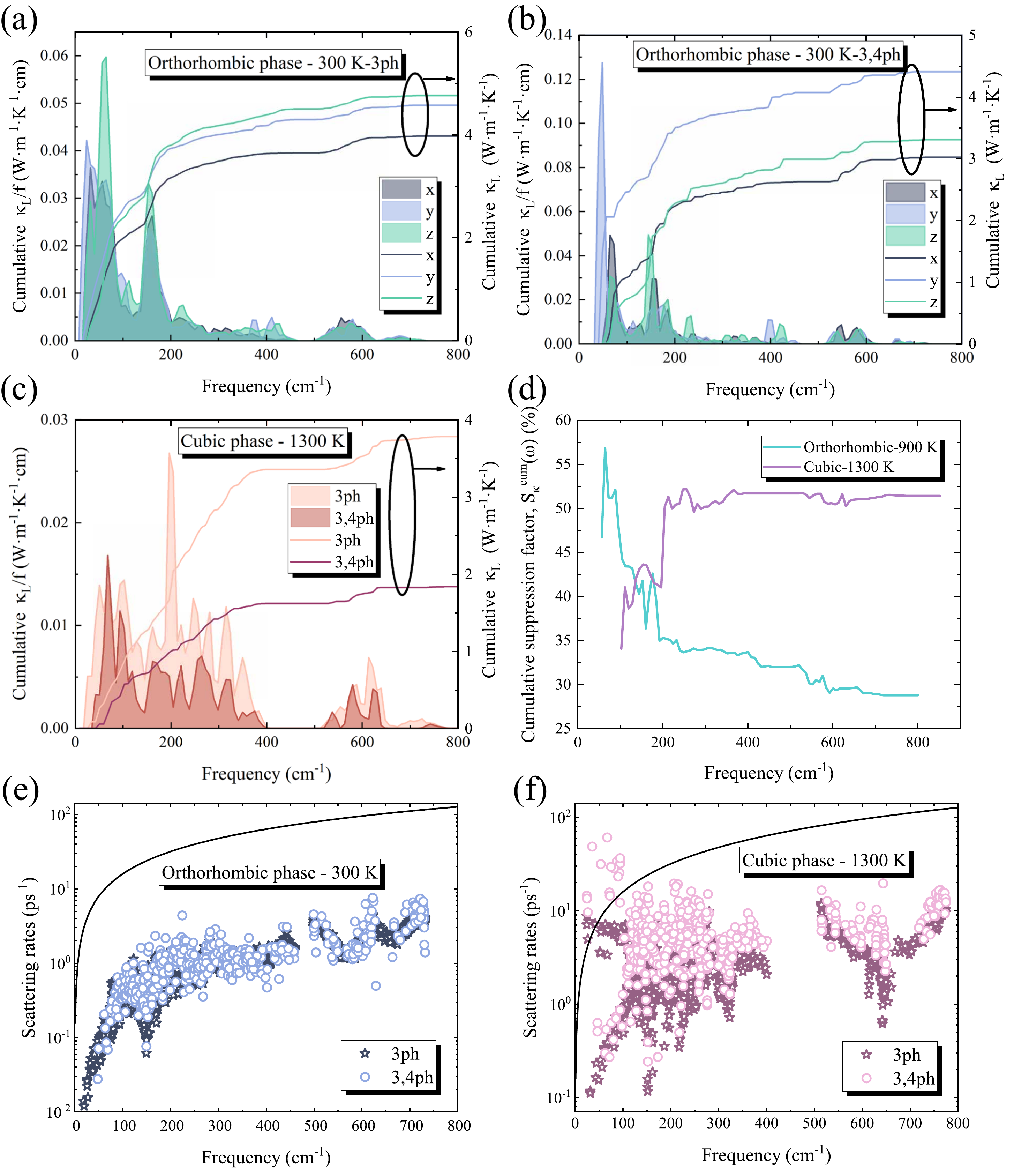}
\caption{
Frequency-resolved phonon transport and scattering properties of SrSnO$_3$. (a,b) Spectral and cumulative particle-like lattice thermal conductivity of the orthorhombic phase at 300~K calculated with three-phonon scattering and with combined three- and four-phonon scattering. (c) Corresponding frequency-resolved thermal conductivity of the cubic phase at 1300~K. (d) Cumulative four-phonon suppression factor \(S_{\kappa}^{\mathrm{cum}}(\omega)\) for the orthorhombic and cubic phases. (e,f) Mode-resolved phonon scattering rates in the orthorhombic phase at 300~K and cubic phase at 1300~K. The solid black line denotes the Ioffe--Regel limit, \(1/\tau=\omega/2\pi\).
}
\label{fig5}
\end{figure*}

Wide-bandgap perovskite oxides are attractive for transparent electronics, ultraviolet optoelectronics, and high-power devices because their corner-sharing $B$O$_6$ networks combine optical transparency, chemical stability, and tunable carrier transport~\cite{Kim2012APEX,Kim2012PRB,Liu2013APL,Ong2015APLMater,Wang2018ACSAMI,Kim2024SciAdv}. Their functional properties are closely coupled to octahedral tilting, lattice instabilities, and anharmonic lattice dynamics, as demonstrated in stannate and titanate perovskites as well as related phase-transforming materials~\cite{Sim2013PRB,Wuttig2018IncipientMetals,Fumega2020SrTiO3,Dangic2021GeTe,Verdi2021Zirconia,He2022SrTiO3Phase}. In addition, recent studies have shown that finite-size effects, structural disorder, and phonon mean-free-path distributions can strongly influence heat transport in perovskite and related oxide systems~\cite{Koh2019AdvSci,Marronnier2018PRB,Zhang2023ACSAMI,Ghosh2025DefectBaSnO3SrSnO3,Zhang2026MTP}. Among alkaline-earth stannates, SrSnO$_3$ stands out as an ultrawide-bandgap oxide with small electron effective masses and favorable n-type transport, making it a candidate for thermally robust electronic applications~\cite{Ong2015APLMater,Wang2018ACSAMI}. In such devices, heat dissipation is ultimately governed by the lattice thermal conductivity ($\kappa_L$), whose magnitude and temperature dependence are set by the frequency distribution of the heat-carrying phonons and the scattering mechanisms that limit their lifetimes~\cite{Broido2007APL,Li2014CPC}.


Recent experiments on epitaxial SrSnO$_3$ films have revealed unusually low $\kappa_L$ and a pronounced thickness dependence, attributed to nanoscale phonon mean free paths and wave-like (coherent) heat transport\cite{Zhang2023APL,Zhang2026MTP}. A complementary theoretical study demonstrated that octahedral tilting softens acoustic phonons and reduces group velocities, establishing structural distortion as an important mechanism for limiting heat transport in this system\cite{Hao2026PRB}. Related perovskites such as SrTiO$_3$ further require finite-temperature phonon renormalization and higher-order anharmonic scattering for accurate thermal transport predictions \cite{Tadano2015PRB,Wang2021PRB,Feng2016PRB,Fumega2020SrTiO3}. Collectively, these studies highlight the central roles of structural distortion and low-frequency phonons in perovskite heat transport\cite{Simoncelli2019NatPhys,PhysRevX.12.041011}. However, the frequency-resolved contributions of four-phonon (4ph) scattering and coherent transport to the renormalized phonon spectrum of SrSnO$_3$ and their quantitative interplay have not been established.


SrSnO$_3$ exists in two experimentally relevant phases: an orthorhombic $Pnma$ structure stable under ambient conditions and a high-symmetry cubic $Pm\bar{3}m$ structure accessible at elevated temperatures~\cite{Zhang2023APL}. The two phases share identical chemistry and atomic mass but differ fundamentally in crystal symmetry and local bonding environment. The orthorhombic phase features corner-connected SnO$_6$ octahedra with cooperative tilting distortions, whereas the cubic phase adopts an undistorted perovskite geometry with higher structural symmetry. This symmetry contrast gives rise to markedly different phonon spectra, group velocities, and scattering phase spaces, making the two phases a controlled pair for isolating how structural distortion and anharmonicity govern thermal transport. By systematically comparing the two phases under identical theoretical treatment, incorporating 4ph scattering, self-consistent phonon renormalization, and coherent transport contributions, we aim to quantify the frequency-resolved mechanisms by which octahedral tilting and higher-order anharmonicity limit $\kappa_L$ in SrSnO$_3$, and 
to clarify how intrinsic anharmonic scattering and coherent transport contribute to the anomalously low $\kappa_L$ observed experimentally.

Here we combine self-consistent phonon renormalization, 3ph and 4ph scattering, and the Wigner coherent contribution to establish a unified microscopic picture of heat transport in SrSnO$_3$. We show that the low-frequency vibrational modes arising from the weak Sr-related sublattice are strongly renormalized with temperature and open efficient higher-order scattering channels. Beyond the previously identified octahedral-tilting-induced acoustic softening, 4ph scattering substantially suppresses the particle-like heat current, reducing the Peierls thermal conductivity by 19.4\% in orthorhombic SrSnO$_3$ at 300 K and by 52.1\% in cubic SrSnO$_3$ at 1300 K. Frequency-resolved analyses show that this suppression occurs over the principal low- and mid-frequency heat-carrying range. At the same time, the Wigner coherent contribution provides a finite compensating channel, so that the apparent agreement between conventional three-phonon calculations and the experimental magnitude of thermal conductivity can be understood as a cancellation between neglected four-phonon scattering and omitted coherent heat transport. These results demonstrate that heat conduction in SrSnO$_3$ cannot be described by a 3ph Peierls picture alone, but requires the combined treatment of phonon renormalization, higher-order anharmonicity, and coherent transport. Detailed information on the software packages, force-constant supercells and cutoff radii, phonon $q$-point meshes, SCPH convergence criteria, and transport-calculation settings is provided in Sec.~V of the Supplementary Information.

Fig.~\ref{fig1} compares the crystal structures and real-space electronic structure of SrSnO$_3$. The orthorhombic phase contains tilted corner-sharing SnO$_6$ octahedra and distorted Sr-centered cages, whereas the cubic phase recovers the untilted $Pm\bar{3}m$ framework. Electron localization and charge-density-difference maps consistently show stronger localization and charge redistribution along Sn-O bonds than around Sr, indicating a mixed ionic–covalent character with a rigid Sn-O backbone embedded in a weakly bonded Sr sublattice.


The orbital-resolved electronic structures and $-$pCOHP curves in Fig.~\ref{fig2}, calculated using LOBSTER~\cite{Lobster2020JCC}, confirm this bonding hierarchy in reciprocal space. The valence bands are dominated by O-$p$ states and the conduction-band edge by Sn-$s$ character, while Sr-projected states contribute weakly near the band edges. The integrated $-$pCOHP values quantify the hierarchy: Sn-O bonding (4.896 eV in the orthorhombic phase, 5.201 eV in the cubic phase) is an order of magnitude stronger than Sr-O (0.301, 0.229 eV) and Sn-Sr (0.438, 0.358 eV) interactions. This stark contrast implies that the weakly constrained Sr sublattice provides low-frequency vibrational channels that are inherently sensitive to anharmonic perturbation.


Fig.~\ref{fig3} shows the temperature-renormalized phonon spectra and projected phonon density of states. In the orthorhombic phase, several low-lying optical branches below 250~cm$^{-1}$ that are associated with Sr displacements and octahedral framework motion shift noticeably with increasing temperature, while the overall spectrum remains dynamically stable up to 1500 K. The projected phonon density of states confirms that Sr vibrations dominate the low-frequency region, Sn modes occupy the intermediate range, and O modes lie at high frequency. The absence of imaginary modes above the orthorhombic stability range indicates local dynamical stability under the imposed symmetry, not thermodynamic preference.


The cubic phase displays stronger renormalization of the low-frequency branches. Soft modes near the zone center and along the $\Gamma$--X--M--$\Gamma$ path are shifted upward as the temperature increases from 1300 to 1600 K, reflecting the stabilization of the cubic lattice by quartic anharmonicity. These renormalized spectra provide the temperature-dependent phonon basis for the transport calculations that follow.

Experimentally, SrSnO$_3$ transforms between two orthorhombic phases near 905 K, to tetragonal $I4/mcm$ near 1062 K, and to cubic $Pm\bar{3}m$ near 1295 K~\cite{Hao2026PRB}. We do not treat the intermediate tetragonal phase. Accordingly, results for orthorhombic or cubic structures outside their equilibrium ranges are symmetry-constrained idealizations used to isolate phase-dependent transport mechanisms, not predictions for the stable phase.

Fig.~\ref{fig4} compares the calculated lattice thermal conductivity with experimental data, while the corresponding values are summarized in Table~\ref{tab:kappa}. In the orthorhombic phase, the three-phonon Peierls contribution decreases rapidly with temperature, following an approximate power-law dependence of $T^{-0.71}$, which is significantly stronger than the experimentally observed trends~\cite{Zhang2023APL,Yasukawa2017CI} of $T^{-0.54\sim -0.56}$. This discrepancy indicates that the three-phonon picture is incomplete in describing the temperature dependence, even though it may appear to improve agreement with the magnitude at a single temperature. 
The remaining difference may arise from effects not included in the present intrinsic calculation, such as finite-temperature lattice expansion, strain, defects, and microstructural scattering in the experimental samples~\cite{doi:10.1021/acs.inorgchem.2c03499,Ghosh2025DefectBaSnO3SrSnO3,Zhang2023APL,Yasukawa2017CI}.

At 300 K, the particle-like thermal conductivity obtained from three-phonon scattering is 4.44 W m$^{-1}$ K$^{-1}$. When four-phonon scattering is included, this value is reduced to 3.58 W m$^{-1}$ K$^{-1}$, corresponding to a suppression of 19.4\%. The coherent contribution is 0.85 W m$^{-1}$ K$^{-1}$, leading to a total lattice thermal conductivity of 4.43 W m$^{-1}$ K$^{-1}$. The apparent agreement between the three-phonon result and the full calculation at room temperature therefore arises from a compensation between four-phonon suppression and the coherent contribution, rather than from a complete physical description.

The full calculation exhibits a weaker temperature dependence of approximately $T^{-0.61}$, which is in better agreement with the experimental scaling than the three-phonon result.

In the cubic phase, four-phonon scattering becomes significantly stronger and constitutes the dominant resistive mechanism. At 1300 K, the particle-like thermal conductivity decreases from 3.84 to 1.84 W m$^{-1}$ K$^{-1}$ when four-phonon processes are included, corresponding to a suppression of 52.1\%. The coherent contribution remains finite at 0.60 W m$^{-1}$ K$^{-1}$, giving a total thermal conductivity of 2.44 W m$^{-1}$ K$^{-1}$. This pronounced suppression reflects the enhanced anharmonic phase space and the increased importance of higher-order scattering in the high-temperature cubic phase.

The experimental samples differ structurally and microstructurally from the ideal bulk crystal considered here. Zhang \textit{et al.} measured a 350-nm-thick single-crystalline epitaxial SrSnO$_3$ film~\cite{Zhang2023APL}, for which finite thickness, the film/substrate interface, epitaxial strain, residual dislocations, and possible oxygen vacancies can modify phonon transport~\cite{Ghosh2025DefectBaSnO3SrSnO3}. Yasukawa \textit{et al.} studied polycrystalline Sr$_{1-x}$La$_x$SnO$_3$ ceramics with grain sizes of several micrometers to approximately 10~$\mu$m~\cite{Yasukawa2017CI}; grain boundaries, porosity, La-induced disorder, and possible oxygen nonstoichiometry provide additional scattering channels. Because quantitative defect concentrations and grain-boundary resistances were not reported, these effects can only be assessed qualitatively. The comparison should therefore be regarded as evidence that the dominant intrinsic mechanisms are captured, rather than as a strict validation of defect-free bulk transport.

The intrinsic bulk value is not a rigorous upper bound for these measurements because the calculated and experimental systems differ in strain, composition, and microstructure. At 300~K, the calculated $\kappa_L$ of 4.43~W~m$^{-1}$~K$^{-1}$ is close to the film value of 4.60~W~m$^{-1}$~K$^{-1}$ and below the ceramic value of approximately 5.4~W~m$^{-1}$~K$^{-1}$~\cite{Zhang2023APL,Yasukawa2017CI}. Epitaxial strain can alter Sn--O bond lengths, octahedral rotations, and the low-frequency heat-carrying modes, partially offsetting boundary and defect scattering~\cite{Wang2018ACSAMI,Ghosh2025DefectBaSnO3SrSnO3}; composition, density, and measurement uncertainties can likewise shift the ceramic and film values. For context, reported uncertainties in related $ABX_3$ perovskite thermal-conductivity measurements are about 7\% for CsPbBr$_3$ films ($0.43\pm0.03$~W~m$^{-1}$~K$^{-1}$) and 19--26\% for MAPb(Br$_x$I$_{1-x}$)$_3$ ($0.27\pm0.07$ to $0.47\pm0.09$~W~m$^{-1}$~K$^{-1}$)~\cite{10.1021/acs.jpclett.9b01053,https://doi.org/10.1002/advs.202401194}. Although these values cannot be transferred quantitatively to SrSnO$_3$, they show that the few-percent crossover in Fig.~\ref{fig4} is comparable to the uncertainty scale encountered in related measurements. Thus, the small crossover does not imply defect-enhanced transport.

The residual difference between the calculated exponent of $-0.61$ and the experimental range of $-0.54$ to $-0.56$ is 0.05--0.07. Weakly temperature-dependent boundary and defect scattering reduces the relative contribution of intrinsic phonon–phonon resistance to the total thermal resistance and therefore tends to flatten the measured $\kappa_L(T)$. Thermal expansion may additionally modify phonon frequencies, group velocities, and anharmonic interactions, although its net influence cannot be determined unambiguously without variable-volume or quasiharmonic calculations. Scattering processes beyond the four-phonon level would introduce additional high-temperature resistance and are therefore unlikely to be the primary origin of the experimentally weaker temperature dependence~\cite{Feng2016PRB,zxcz-5q9w}. We consequently attribute the remaining deviation mainly to boundary and defect scattering, epitaxial strain, sample-dependent composition and microstructure, and experimental uncertainty, with thermal expansion providing a possible additional correction.

The frequency-resolved results in Fig.~\ref{fig5} distinguish the present mechanism from a purely acoustic-softening picture by identifying directly which heat-carrying phonons are suppressed by higher-order anharmonicity. For orthorhombic SrSnO$_3$ at 300 K, the cumulative particle-like thermal conductivity increases primarily below approximately 250 cm$^{-1}$, a frequency window dominated by acoustic branches and low-lying optical modes, while high-frequency O-dominated modes contribute only marginally. Inclusion of four-phonon scattering reduces the spectral weight within this same low-frequency window, indicating that four-phonon processes act predominantly on the main heat-carrying phonon modes rather than on the high-frequency tail.

For cubic SrSnO$_3$ at 1300 K, the heat-carrying contribution extends over a broader low- and intermediate-frequency range. We quantify the four-phonon-induced reduction through the cumulative suppression factor,
based on the frequency-resolved cumulative thermal conductivity $\kappa_{\mathrm{cum}}(\omega)$ commonly used in spectral and mean-free-path–resolved transport analyses~\cite{Minnich2015PRL,regner2013broadband,chen2021non}:
\begin{equation}
S_{\kappa}^{\mathrm{cum}}(\omega)=
\left[1-\frac{\kappa_{\mathrm{cum}}^{3,4\mathrm{ph}}(\omega)}{\kappa_{\mathrm{cum}}^{3\mathrm{ph}}(\omega)}\right]\times100\% .
\end{equation}

This definition provides a frequency-resolved measure of how four-phonon scattering suppresses particle-like heat-carrying channels relative to the three-phonon baseline and is widely used in spectral phonon transport analyses to resolve the contribution of different scattering mechanisms to lattice thermal conductivity.

This factor reaches approximately 19.4\% in the orthorhombic phase and increases to about 52.1\% in the cubic phase, demonstrating that four-phonon scattering affects the dominant heat-carrying frequency window in both phases, with a substantially stronger effect in the high-temperature cubic phase.

The mode-resolved scattering rates further support this picture: most orthorhombic modes remain well below the Ioffe--Regel limit and retain well-defined quasiparticle character~\cite{wang2023role,wang2024anomalous,wang2024revisiting,PhysRevX.12.041011}, whereas in the cubic phase several modes approach this limit once four-phonon scattering is included, reflecting the enhanced breakdown of the purely particle-like description.

Importantly, this does not imply a complete dominance of coherent transport: the coherent contribution remains finite and slowly varying, providing a secondary channel that does not overtake particle-like transport, as shown in Fig.~\ref{fig4}. Phase-space analysis further confirms the origin of the enhanced scattering, revealing additional anharmonic decay channels opened by quartic interactions within the main heat-carrying frequency window.

Taken together, these results identify four-phonon scattering as the decisive mechanism limiting particle-like heat transport in SrSnO$_3$, with octahedral-tilting-induced acoustic softening providing the structural background and a finite coherent contribution required to recover the experimentally relevant magnitude of $\kappa_L$.

In conclusion, we have developed a unified first-principles framework for lattice thermal transport in orthorhombic and cubic SrSnO$_3$, combining phonon renormalization, three- and four-phonon scattering, and coherent transport. The results reveal a consistent picture in which structural distortion and anharmonicity must be evaluated together of perovskite heat conduction.

Four-phonon scattering is the dominant mechanism suppressing particle-like transport, while a finite coherent contribution provides an additional wave-like channel.

The apparent agreement between three-phonon calculations and experiment at 300 K is accidental, arising from a cancellation between four-phonon suppression and coherent transport. Including both effects yields $\kappa_{\rm p}^{3,4ph} + \kappa_{\rm c}$, which improves both magnitude and temperature dependence, bringing the scaling closer to experiment ($T^{-0.61}$ vs $T^{-0.54\sim-0.56}$).

More generally, reliable prediction of thermal transport in strongly anharmonic oxides requires explicit inclusion of both four-phonon and coherent contributions.

\section*{ACKNOWLEDGMENTS}
%
%
We acknowledge the support from the National Natural Science Foundation of China 
(No.52595632).  
This work was sponsored by the Key Research and Development Program of China (No.2023YFB4604100). 
Additional support was provided by the Xi’an Jiaotong University Doctoral Interdisciplinary Cultivation Program (No. IDT2208).
%
We also acknowledge the support by HPC Platform, Xi’an Jiaotong University. 

\section*{SUPPLEMENTARY MATERIAL}
See the Supplementary Material for Validation of the Calculated Structure and Temperature-Renormalized Phonon Spectra, Characteristic-Length Dependence of the Population Thermal Conductivity, additional temperature-dependent phonon group velocities, phonon scattering rates including three- and four-phonon processes, channel-resolved anharmonic scattering phase spaces, frequency-resolved phase-space analyses, and further computational details.

\bibliography{References}

\end{document}